\documentclass[twocolumn, 10pt]{autart}
\usepackage{enumerate}
\usepackage{graphicx}
\usepackage{amsmath, amssymb}
\usepackage{epstopdf} 
\usepackage{mathtools}
\usepackage{booktabs}
\usepackage[american]{babel}
\usepackage{xcolor, soul}
\usepackage{siunitx}    
\usepackage{mathabx}
\usepackage{accents}
\usepackage{textcomp}
\usepackage{wrapfig}
\usepackage{hyperref}
\usepackage{url}

\usepackage[firstpageonly=true]{draftwatermark}	

\DeclareMathOperator{\Int}{Int}

\newtheorem{theorem}{Theorem}

\newtheorem{assumption}{Assumption}
\newtheorem{proposition}{Proposition}
\theoremstyle{definition}
\newtheorem{definition}{Definition}
\newtheorem{remark}{Remark}
\newtheorem{problem}{Problem}

\newcommand\scalemath[2]{\scalebox{#1}{\mbox{\ensuremath{\displaystyle #2}}}}

\newcommand\ubar[1]{\underaccent{\bar}{#1}}

\hypersetup{
	colorlinks,
	linkcolor={blue!50!black},
	citecolor={blue!50!black},
	urlcolor={blue!80!black}
}
  
\edef\endfrontmatter{
  \unexpanded\expandafter{\endfrontmatter}
  \noexpand\endNoHyper 
}

\begin{document}

\begin{frontmatter}

\title{Nonlinear MPC design for incrementally ISS systems with application to GRU networks}

\thanks[footnoteinfo]{Corresponding author.}
\thanks[uuse]{\,Fabio Bonassi is now with the Department of Information Technology of the Uppsala University, Uppsala, Sweden.}

\author[polimi]{Fabio Bonassi\thanksref{footnoteinfo}\thanksref{uuse}},
\author[polimi]{Alessio La Bella},
\author[polimi]{Marcello Farina},
\author[polimi]{Riccardo Scattolini}

\address[polimi]{Dipartimento di Elettronica, Informazione e Bioingegneria, Politecnico di Milano, Via Ponzio 34/5, 20133, Milano, Italy. E-mail: {\tt name.surname@polimi.it}}


\begin{abstract}
This brief addresses the design of a Nonlinear Model Predictive Control (NMPC) strategy for exponentially incremental Input-to-State Stable (ISS) systems. In particular, a novel formulation is devised, which does not necessitate the onerous computation of terminal ingredients, but rather relies on the explicit definition of a minimum prediction horizon ensuring closed-loop stability. 
The designed methodology is particularly suited for the control of systems learned by Recurrent Neural Networks (RNNs), which are known for their enhanced modeling capabilities and for which the incremental ISS properties can be studied thanks to simple algebraic conditions.
The approach is applied to Gated Recurrent Unit (GRU) networks, providing also a method for the design of a tailored state observer with convergence guarantees. 
The resulting control architecture is tested on a benchmark system, demonstrating its good control performances and efficient applicability.
\end{abstract}

\end{frontmatter}

\DraftwatermarkOptions{%
 angle=0,
 hpos=104mm,
 vpos=270mm,
 fontsize=2.5mm,
 color={[gray]{0.2}},
 text={
   \parbox{0.99\textwidth}{© 2023. This manuscript version is made available under the CC-BY-NC-ND 4.0 license \url{https://creativecommons.org/licenses/by-nc-nd/4.0/} \\
   This manuscript has been accepted for publication at Elsevier Automatica. Please cite the published article instead of this manuscript. \\
   DOI: \href{https://doi.org/10.1016/j.automatica.2023.111381}{10.1016/j.automatica.2023.111381}
   }}}


\section{Introduction}
Model Predictive Control (MPC) is today a well-established control strategy, largely studied from the methodological point of view and widely adopted in many engineering applications~\cite{rawlings2017model}. 
Given the necessity of a dynamical representation of the system under control, not always available in practice, a huge interest has risen on the design of learning-based MPC regulators relying on black-box models derived from data~\cite{hewing2020learning}. In this context, Recurrent Neural Networks (RNNs) have recently gained increasing interest, showing to be particularly suited for representing nonlinear dynamical systems~\cite{mohajerin2019multistep}. 
This motivated the wide use of RNNs for predictive control, e.g., in process industry~\cite{lanzetti2019recurrent}, and in chemical applications~\cite{wu2020process}.

Despite their potentialities, few theoretical results have been established on the stability properties of RNN-based control systems. 
Considering RNNs in open-loop configurations, sufficient conditions to enforce the exponential \textit{Incremental Input-to-State Stability} ($\delta$ISS) during their training process have been derived in~\cite{bonassi2022survey}. 
This property plays a crucial role in \mbox{RNNs}, ensuring that state trajectories asymptotically depend solely on the applied inputs, implying that modeling performances are independent on the network state initialization.
On the other hand, guaranteeing the closed-loop stability of \mbox{RNN-based} MPC regulators is still an open challenge, given the heterogeneity of RNN architectures and their high nonlinearity.
In fact, classic methods to attain closed-loop stability of Nonlinear MPC (NMPC) regulators rely on the definition of additional ingredients, such as a terminal cost function, a terminal constraint set, and a local stabilizing control law~\cite{mayne2000constrained}. These ingredients can be onerous to compute~\cite{magni2001stabilizing} and, moreover, they may need to be online redefined whenever the closed-loop system reference changes~\cite{kohler2019nonlinear}. 

In the context of designing NMPC laws with stability guarantees for RNN models, several approaches have been proposed, albeit tailored to specific classes of RNNs.
For example, \cite{terzi2021lstm} and \cite{armenio2019model} propose a NMPC strategy for Long Short-term Memory (LSTM) network and Echo State Network (ESN) models, respectively, by exploiting their specific structure for the definition of suitable terminal costs.
The design of a closed-loop stable NMPC regulator for Neural Network AutoRegressive eXogenous (NNARX) architectures with one hidden layer is presented in \cite{patan2014neural}. 
The same RNN class {in a multi-layer framework} is considered in \cite{bonassi2022offset}, where a zero-terminal constraint is imposed to ensure closed-loop NMPC stability, avoiding the online computation of the terminal constraint set at the price of more conservative performances.  On the other hand, the NMPC design for Gated Recurrent Units (GRUs) networks is presented in \cite{bonassi2021nonlinear}, however involving the online computation of terminal ingredients and assuming the existence of an auxiliary control law. For continuous-time single-layer RNN architectures, a stabilizing MPC strategy is presented in \cite{wu2019machine}, where the terminal ingredients are replaced by explicit constraints on a Lyapunov function.

In view on the fact that existing RNN-based NMPC solutions with stability guarantees are tailored to specific network architectures, the present work proposes a closed-loop stable NMPC strategy that can be applied to any generic nonlinear system, provided that it features exponential $\delta$ISS.
This property can be effectively enforced to notable RNN architectures by adding suitable constraints to the training problem, as shown in \cite{bonassi2022survey}. 
The proposed strategy ensures closed-loop stability without requiring the definition of terminal ingredients, as in classic NMPC methods \cite{mayne2000constrained}, thus yielding a more efficient NMPC design procedure.
In addition to the general formulation of the proposed strategy, we also discuss  how this can be applied to the GRU networks, characterized by advanced modeling performances and efficient training procedure, as shown in  
\cite{zarzycki2021lstm}. 
To this end, we propose a methodology for designing a nominally convergent observer for black-box GRU models. 

The paper is structured as follows. The control problem is formulated in Section \ref{sec:control}, and the proposed NMPC strategy is described in Section \ref{sec:mpc}. Section \ref{sec:gru} presents the synthesis of the control architecture for GRU networks, and the design of a suitable state observer. The approach is tested on a referenced chemical process in Section \ref{sec:casestudy}. Conclusive considerations are derived in Section \ref{sec:conclusion}.

\smallskip
\subsection*{Notation}
Given a vector $v \in \mathbb{R}^n$, we denote by $v^\prime$ its transpose, by $[ v ]_i$ its $i$-th component, and by $\| v \|_p$ its $p$-norm.
Moreover, given a square matrix $A$, $\| v \|_A^2$ is used to indicate the quadratic form $v^\prime A v$.
We denote the Hadamard product between two vectors $v$ and $w$ as $v \circ w$. 
The time index $k$ of time-varying vectors is reported as a subscript, e.g. $v_k$. 
Sequences of vectors spanning from the time index $k_1$ to $k_2 \geq k_1$ are indicated by $v_{k_1 : k_2}$, i.e. $v_{k_1 : k_2} = \{ v_{k_1}, v_{k_1 +1}, ..., v_{k_2} \}$.
The $\ell_{p, q}$ norm of a sequence is defined as 
\mbox{$	\| v_{k_1 : k_2} \|_{p, q} = \big\| \, [ \| v_{k_1} \|_p,  \| v_{k_1 + 1} \|_p, ..., \| v_{k_2} \|_p ]^\prime \, \big\|_q. $}
A notable case is the $\ell_{p, \infty}$ norm, for which $\| v_{k_1 : k_2} \|_{p, \infty} = \max_{k \in \{ k_1, ..., k_2\} } \| v_k \|_p$. 
Given a matrix $A$, $ \| A \|_p$ is used to indicate its induced $p$-norm, whereas $\bar{\varsigma}_A$ and $\ubar{\varsigma}_A$ denote its maximum and minimum singular values, respectively.
Finally, we denote by $\sigma(x) = \frac{1}{1 + e^{-x}}$  and by $\phi(x) = \tanh(x)$ the sigmoidal  and $\tanh$ activation functions, respectively.
Note that for vector arguments, these functions are intended to be applied element-wise.

\section{Control problem} \label{sec:control}
{Consider a discrete-time nonlinear system described by}
\begin{equation} \label{eq:control:model}
	\Sigma: \begin{dcases}
		x_{k+1} = \varphi(x_k, u_k) \\
		y_k = g(x_k)
	\end{dcases},
\end{equation}
where $u_k \in \mathbb{R}^{n_u}$, $x_k \in \mathbb{R}^{n_x}$, and $y_k \in \mathbb{R}^{n_y}$ are the input, state, and output vector, respectively.
It is assumed that the input is constrained {in a compact set $\mathcal{U} \subset \mathbb{R}^{n_u}$}, and that the system \eqref{eq:control:model} admits an invariant set $\mathcal{X}\subseteq \mathbb{R}^{n_x}$, that is, \mbox{$x \in \mathcal{X} \implies \varphi(x, u) \in \mathcal{X} $} for any $u \in \mathcal{U}$.

The generic system \eqref{eq:control:model} can represent many different RNN architectures \cite{bonassi2022survey}. Note that, in that case, the definition of the invariant set $\mathcal{X}$ is related to the considered RNN class, see, e.g., \cite{terzi2021lstm}, \cite{bonassi2021stability}, \cite{armenio2019echo}, where an invariant set is proposed for LSTMs, GRUs, and ESNs, respectively.
It is furthermore assumed that system \eqref{eq:control:model} is exponentially $\delta$ISS according to the following definition.

\smallskip
\begin{definition}[Exponential $\delta$ISS]
	System \eqref{eq:control:model} is exponentially $\delta$ISS if there exist constants $\mu > 0$ and $\lambda \in (0, 1)$, and a $\mathcal{K}_\infty$ function $\gamma$, such that, for any pair of initial states $x_{a, 0} \in \mathcal{X}$ and $x_{b, 0} \in \mathcal{X}$ and any pair of input sequences $u_{a, 0:k} \in \mathcal{U}_{0:k}$ and $u_{b, 0:k} \in \mathcal{U}_{0:k}$, it holds that
	\begin{equation} \label{eq:deltaiss:def}
		\| x_{a, k} - x_{b, k} \|_2 \leq \mu \lambda^k \| x_{a, 0} - x_{b, 0} \|_2 + \gamma(\| u_{a, 0:k} - u_{b, 0:k} \|_{2, \infty}),
	\end{equation}
	where $x_{\alpha, k} = x_{\alpha, k}( x_{\alpha, 0},  u_{\alpha, 0:k})$ denotes the state trajectory of \eqref{eq:control:model}, initialized in $ x_{\alpha, 0}$ and fed with the input $u_{\alpha, 0:k}$, for $\alpha \in \{ a, b \}$.
\end{definition}

This stability property implies that, within the invariant set $\mathcal{X}$, the effect of initial conditions on state trajectories asymptotically vanishes, and that the closer two input sequences are, the smaller is the maximum $\ell_2$ distance between the resulting state trajectories.
This rather strong stability property has been investigated for RNN models \cite{bonassi2022survey}, and theoretically-sound training strategies to obtain exponentially $\delta$ISS RNN models have been proposed, see \cite{terzi2021lstm} \cite{bonassi2021stability}, and \cite{armenio2019echo}.
Notably, these works allow to conservatively compute values $\mu$ and $\lambda$, as function of the network's weights, that satisfy \eqref{eq:deltaiss:def}.
This computation is illustrated, for GRU networks, in Section \ref{sec:gru}: we here assume that these values exist and can be computed.

\smallskip
\subsection{Problem statement}
Given the exponentially $\delta$ISS system~\eqref{eq:control:model}, an output-feedback control architecture is designed.
The control goal is to steer the plant's output to a (piece-wise) constant setpoint $\bar{y}$, while fulfilling the input constraint $u_k \in \tilde{\mathcal{U}}$, where $\tilde{\mathcal{U}} \subseteq \mathcal{U}$ denotes a compact set, potentially more tight than $\mathcal{U}$.
To address this regulation problem, it is first necessary to assume the existence of a feasible equilibrium corresponding to the output setpoint. 

\medskip
\begin{assumption} \label{ass:control:equilibrium}
	Given the output reference $\bar{y}$, there exists $\bar{x} \in \Int(\mathcal{X})$ and $\bar{u} \in \Int(\tilde{\mathcal{U}})$ such that the triplet $\bar{\Sigma} = (\bar{x}, \bar{u}, \bar{y})$ is a feasible equilibrium, i.e. such that $\bar{x} = \varphi(\bar{x}, \bar{u})$ and $\bar{y} = g(\bar{x})$.
\end{assumption} 

Therefore, considering a setpoint $\bar{y}$ that fulfills Assumption \ref{ass:control:equilibrium}, the control problem can be stated as follows.

\medskip
\begin{problem}(Regulation problem) \label{problem:control}
	Given the model \eqref{eq:control:model} and the output setpoint $\bar{y}$, steer the system to the equilibrium $\bar{\Sigma}$ by means of a control action that satisfies the input constraint $u_k \in \tilde{\mathcal{U}}$.
\end{problem}

\subsection{Proposed control architecture}

\begin{figure}[t]
	\centering
	\includegraphics[width=0.9 \linewidth]{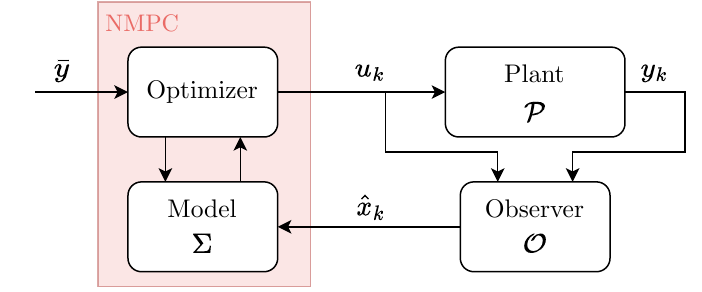}
	\caption{Scheme of the considered control architecture.}
	\label{fig:control_architecture}
\end{figure}

We propose to address Problem \ref{problem:control} by means of the NMPC architecture depicted in Figure \ref{fig:control_architecture}, which requires (\emph{i}) the design of an exponentially converging state observer, and (\emph{ii}) the synthesis of a state-feedback NMPC control law.

Considering a generic state observer,
\begin{equation} \label{eq:control:generic_observer}
	\mathcal{O}: \begin{dcases}
		\hat{x}_{k+1} = \varphi_o(\hat{x}_k, u_k, y_k) \\
		\hat{y}_k = g_o(\hat{x}_k)
	\end{dcases},
\end{equation}
the definition of {weak exponential observer} follows.

\medskip
\begin{definition}[{Weak exponential observer \cite{magni2004stabilization}}] \label{def:observer_convergence}
    The observer \eqref{eq:control:generic_observer} of system \eqref{eq:control:model} is said to be a weak exponential observer if there exist constants $\mu_o > 0$ and $\tilde{\lambda}_o \in (0, 1)$ such that, for any initial state $x_0 \in \mathcal{X}$, given the sequence of applied input $u_{0:k} \in \mathcal{U}$ and the sequence of measured output $y_{0:k}(x_0, u_{0:k})$, it holds that
	\begin{equation}  \label{eq:control:observer_convergence}
    	\| \hat{x}_k(\hat{x}_0, u_{0:k}, y_{0:k}) - x_k(x_0, u_{0:k}) \|_2 \leq \mu_o \tilde{\lambda}_o^k \| \hat{x}_0 - x_0 \|_2.
	\end{equation}
\end{definition}

The design of observer satisfying Definition \ref{def:observer_convergence} is not restrictive for the considered RNN-based models.
Indeed, observers with convergence guarantees have been proposed in \cite{armenio2019echo} and \cite{terzi2021lstm} for ESNs and LSTMs models, respectively.
The synthesis procedures described in the above references rely upon the models' $\delta$ISS to design and optimally tune the state observers.
For the design of state observers for GRUs the reader is addressed to Section \ref{sec:gru}, where a novel synthesis technique is devised.

The second of the control architecture of Figure \ref{fig:control_architecture} is a state-feedback NMPC law \cite{rawlings2017model}.
{To this end, 
we propose an NMPC law that addresses Problem~\ref{problem:control}, and we provide explicit sufficient conditions on its design parameters that guarantee the closed-loop stability of the scheme.}

\section{Proposed Nonlinear MPC formulation} \label{sec:mpc}
According to the predictive control paradigm \cite{rawlings2017model}, the proposed NMPC law is defined by formulating the underlying Finite Horizon Optimal Control Problem (FHOCP) to be solved at every discrete time step $k$, based on the predictive model of the system \eqref{eq:control:model}, allowing to predict the future state trajectories throughout the prediction horizon $N$, given the current state estimate $\hat{x}_k$ yielded by the observer \eqref{eq:control:generic_observer} and the applied input.

For the sake of clarity, at time $k$, we denote by $u_{k:k+N-1 \lvert k} = \{ u_{k \lvert k}, ..., u_{k+N - 1 \lvert k}  \}$
the sequence of future inputs applied throughout the prediction horizon $\mathcal{N} = \{ 0, ..., N - 1 \}$, and by $x_{k:k+N \lvert k} = \{ x_{k \lvert k}, ..., x_{k+N \lvert k}  \}$ the predicted state trajectory, where $x_{k \lvert k} = \hat{x}_k$.
Note that the term $x_{k + t \lvert k}$ denotes the predicted state at time $k+t$ given the input sequence $u_{k:k+t-1 \lvert k}$. 
Under this notation, the FHOCP can therefore be stated as follows.
\begin{subequations} \label{eq:mpc:fhocp}
	\begin{align}
		\min_{u_{k:k+N_c-1 \lvert k}} & \Big \{ J_k = \sum_{\tau=0}^{N - 1} \big( \| x_{k+\tau \lvert k} - \bar{x} \|_Q^2  \label{eq:mpc:fhocp:cost} \\
        & \qquad+ \| u_{k+\tau \lvert k} - \bar{u} \|_R^2  \big)  + V_{\bar{\Sigma}}(x_{k+N \lvert k}) \Big \} \nonumber\\
		\text{s.t.} \quad & x_{k \lvert k} = \hat{x}_k \label{eq:mpc:fhocp:initialization} \\
		& x_{k+\tau +1 \lvert k} = \varphi(x_{k+\tau \lvert k}, u_{k + \tau \lvert k} ) \quad\!\! \forall \tau \in \mathcal{N} \label{eq:mpc:fhocp:dynamics} \\
		& u_{k + \tau \lvert k} \in \tilde{\mathcal{U}} \qquad\qquad\qquad\qquad\quad\!\! \forall \tau \in \mathcal{N} \label{eq:mpc:fhocp:input}
	\end{align}
In the above formulation, the dynamics of the predictive model is embedded by means of the constraints \eqref{eq:mpc:fhocp:dynamics}.
With constraint \eqref{eq:mpc:fhocp:initialization}, the predictive model is initialized at the state estimate $\hat{x}_k$ yielded by the observer \eqref{eq:control:generic_observer}.
The input constraint is enforced via \eqref{eq:mpc:fhocp:input}.
The quadratic cost function $J_k$ in \eqref{eq:mpc:fhocp:cost} penalizes the deviation (throughout the horizon) of the predicted state trajectory from $\bar{x}$, and the deviation of the input sequence from $\bar{u}$.
These terms are weighted by the positive definite matrices $Q$ and $R$, respectively.
The term {$V_{\bar{\Sigma}}(x_{k+N \lvert k})$} is the terminal cost, and is commonly designed to approximate the cost-to-go from $x_{k+N \lvert k}$ to the target equilibrium $\bar{x}$.
Albeit designing such term is generally onerous~\cite{magni2001stabilizing}, the exponential $\delta$ISS of model \eqref{eq:control:model} is here leveraged to retrieve a general formulation of $V_{\bar{\Sigma}}(x_{k+N \lvert k})$ which ensures that the MPC law is closed-loop stable.
In particular, we consider a terminal cost defined as
\begin{equation}
	V_{\bar{\Sigma}}(x_{k+N \lvert k}) = \sum_{\tau= 0}^{M} \big\| x_{k+N+\tau \lvert k} - \bar{x} \big\|_S^2,
\end{equation}
where $x_{k+N+\tau \lvert k}$, with $\tau \in \{ 0, ..., M \}$, denotes the evolution of \eqref{eq:control:model}, from the end of the prediction horizon (i.e., $k+N$) for $M$ steps in the future, under the constant input $\bar{u}$. 
That is, \mbox{$x_{k+N+\tau +1 \lvert k} = \varphi(x_{k+N+\tau \lvert k}, \bar{u} )$}\,,\linebreak  \mbox{$\forall \tau \in \{ 0, \hdots, M - 1\}.$}
Explicit sufficient conditions on the choice of the positive scalar $M$, named simulation horizon, and of the weight matrix $S$ are detailed later. 
\end{subequations}

According to the receding horizon principle, at time $k$ the optimal control sequence $u_{k:k+N-1 \lvert k}^\star$ is computed by solving the FHOCP \eqref{eq:mpc:fhocp}, and the fist optimal control input $u_k = u_{k \lvert k}^\star$. 
This implicit procedure yields the NMPC control law \mbox{$u_k = \kappa_{\textrm{MPC}}(\hat{x}_k)$}.

\medskip
\begin{theorem} \label{theorem:mpc_convergence}
	A sufficient condition for the closed-loop stability of the NMPC law $u_k = \kappa_{\textrm{MPC}}(\hat{x}_k)$ associated to the FHOCP  \eqref{eq:mpc:fhocp} is that the weight matrices $Q$ and $S$ are designed so that
	\begin{subequations} \label{eq:mpc:as_conditions}
\begin{equation}\label{eq:mpc:sigmas_condition}
		\bar{\varsigma}_Q < \ubar{\varsigma}_S
	\end{equation}
	and that the simulation horizon $M$ satisfies
	\begin{equation} \label{eq:mpc:horizon_condition}
		M > \frac{1}{2} \log_{\lambda} \left( \frac{\ubar{\varsigma}_S - \bar{\varsigma}_Q}{\mu^{2}  \bar{\varsigma}_S} \right) - 1.
	\end{equation}
	\end{subequations}
\end{theorem}
\begin{pf}
	See Appendix \ref{proof:mpc_convergence}.
\end{pf}

It is worth to highlight that the design of the terminal cost is arbitrary, as the weight matrix $S$ is chosen freely (subject to \eqref{eq:mpc:sigmas_condition}).
Secondly, \eqref{eq:mpc:horizon_condition} represents an explicit condition on the simulation horizon of the system beyond the prediction horizon. 

\smallskip
\begin{remark} The existence of an  $M$ large enough to guarantee closed-loop NMPC stability has been considered in \cite{magni2001stabilizing} and \cite{soloperto2022nonlinear}, but an explicit lower bound was not provided. 
In fact, the attainment of nominal closed-loop stability guarantees via a suitably long  horizon has also been discussed in \cite{boccia2014stability}, where it has been proven that terminal cost and terminal constraint could be removed if quasi-infinite horizons are adopted.
	Seen through these lenses, the proposed NMPC formulation can be seen as a quasi-infinite horizon formulation, with control horizon $N$ and prediction horizon $N+M$, where the minimum provenly-stabilizing prediction horizon is now explicitly known.
\end{remark}

\section{Application to  GRUs} \label{sec:gru}
In order to illustrate how the proposed control architecture can be practically adopted for RNNs enjoying the $\delta$ISS property, we consider below the  case of Gated Recurrent Units (GRUs).
For this architecture, in the following we describe how to synthesize the control elements discussed in the previous sections, namely the exponentially converging state observer and the state-feedback NMPC law guaranteeing the asymptotic stability of the closed-loop, to address Problem \ref{problem:control}.

Finally, let us reiterate that the proposed strategy can also be applied to all other RNN architectures, for which $\delta$ISS guarantees are available \cite{bonassi2022survey}. 
One might thus consider a $\delta$ISS LSTM alongside with the converging state observer proposed in \cite{terzi2021lstm}.

\subsection{Gated Recurrent Units}
Consider the following nonlinear discrete-time state-space system, describing a single-layer GRU model,
\begin{subequations} \label{eq:gru:model}
\begin{equation} \label{eq:gru:model:state_space}
	\Sigma(\Phi): \, \begin{dcases}
		x_{k+1} = z_k \circ x_k + (1 - z_k) \circ r_k \\
		y_k = U_o \, x_k + b_o
	\end{dcases}.
\end{equation}
The vector $x_k \in \mathbb{R}^{n_x}$ represents the GRU's state, $u_k \in \mathbb{R}^{n_u}$ its input vector, and $y_k \in \mathbb{R}^{n_y}$ its output.
Note that the state dimensionality $n_x$  matches the number of units of the layer, which is a design choice of the model.
The term $r_k$ is the so-called \emph{squashed input}, which reads as 
\begin{equation}\label{eq:squashedinput}
	r_k = \phi( W_r \, u_k + U_r \, f_k \circ x + b_r ).
\end{equation} The terms $z_k$ in \eqref{eq:gru:model:state_space} and $f_k$ in \eqref{eq:squashedinput} are named \emph{update} and \emph{forget} gates, respectively.
These gates are functions of the inputs and states, squashed by the sigmoidal activation function:
\begin{equation} \label{eq:gru:model:gates}
	\begin{aligned}
		z_k &= \sigma( W_z u_k + U_z x_k + b_z), \\
		f_k &= \sigma( W_f u_k + U_f x_k + b_f).
	\end{aligned}
\end{equation}
\end{subequations} 
It is worth noticing that the shallow GRU model \eqref{eq:gru:model} falls into the general form \eqref{eq:control:model} and that it depends on the set of weights $\Phi = \{ W_z, U_z, b_z, W_f, U_f, b_f, W_r, U_r, b_r, U_o, b_o \}$.
These weights parametrize the model and need to be tuned during the so-called training procedure.

At this point, it is worth recalling from \cite{bonassi2021stability} the regional stability properties enjoyed by shallow GRUs.
To this end, it is first necessary to define an invariant set with respect to which such properties are stated.
The following customary assumptions concerning the boundedness of the model's input and of the model's initial state candidates are required.

\smallskip
\begin{assumption} \label{ass:bounded_input}
	The input of model \eqref{eq:gru:model} is unity-bounded, i.e.,
	\begin{equation} \label{eq:gru:input_set}
		u_k \in \mathcal{U} = \{ u \in \mathbb{R}^{n_u} : \| u \|_\infty \leq 1 \}.
	\end{equation}
\end{assumption}

\smallskip
\begin{assumption} \label{ass:bounded_initial_set}
	The initial state of the shallow GRU \eqref{eq:gru:model} belongs to an arbitrarily large, but finite, set
	\begin{equation} \label{eq:gru:initial_set}
		\mathcal{X} = \{ x \in \mathbb{R}^{n_x} : \| x \|_\infty \leq \check{x} \},
	\end{equation}
	with $\check{x} \geq 1$.
\end{assumption}

Assumption \ref{ass:bounded_input} is customary when working with neural networks, see \cite{goodfellow2016deep}, and can be easily satisfied by means of normalization procedures as long as the input is saturated.
In closed-loop operation, this is ensured by constraining the control variable, see \eqref{eq:mpc:fhocp:input}.
Concerning Assumption \ref{ass:bounded_initial_set}, in  \cite{bonassi2021stability} it has been shown that \eqref{eq:gru:initial_set} represents an invariant set of the GRU.

In this context, the stability properties of \eqref{eq:gru:model:gates} have been analyzed in \cite{bonassi2021stability}, where a sufficient condition on the weights $\Phi$ guaranteeing the GRU's exponential $\delta$ISS has been devised. 
Specifically, this condition consists of a nonlinear inequality on $\Phi$, which can be readily enforced during the training procedure to ensure that the trained network enjoys the $\delta$ISS property.
Relying on these existing results, we henceforth assume that \eqref{eq:gru:model} has been trained accordingly, so that it enjoys the exponential $\delta$ISS property in virtue of \cite[Theorem 2]{bonassi2021stability}.
Note that such theorem not only guarantees the existence of the exponential $\delta$ISS-related constants, i.e. $\mu$ and $\lambda$, but it can be exploited to provide an explicit estimate of such values, by means of the following proposition.

\smallskip
\begin{proposition} \label{prop:exp_deltaiss}
	Given the exponentially $\delta$ISS GRU \eqref{eq:gru:model}, a conservative estimate of $\mu$ and $\lambda$ of \eqref{eq:deltaiss:def}  is
    \begin{subequations} \label{eq:gru:exp_deltaiss:mu_lambda}
        \begin{align}
            \mu  &= \sqrt{n_x}, \\
            \lambda &= \max(\kappa_x(\check{\sigma}_z), \kappa_x(1-\check{\sigma}_z)), \label{eq:gru:exp_deltaiss:mu_lambda:lambda}
        \end{align}
    \end{subequations}
    where the function $\kappa_x(\cdot)$ is defined as
    \begin{equation} \label{eq:gru:exp_deltaiss:kappa}
    \scalemath{0.9}{
    \begin{aligned}
        \kappa_x(z) =& z + (1-z) \Big( \frac{1}{4} \check{x} \| U_f \|_\infty + \check{\sigma}_f \Big) \| U_r \|_\infty+ \frac{1}{4} (\check{\phi}_r + \check{x}) \| U_z \|_\infty,
    \end{aligned}}
    \end{equation}
    and the terms $\check{\sigma}_f$, $\check{\sigma}_z$, and $\check{\phi}_r$ read as
    \begin{subequations} \label{eq:gru:exp_deltaiss:gates}
        \begin{align}
            \check{\sigma}_f = \sigma( \| W_f \quad U_f \, \check{x} \quad b_f \|_\infty), \\
            \check{\sigma}_z = \sigma( \| W_z \quad U_z \, \check{x} \quad b_z \|_\infty), \label{eq:gru:exp_deltaiss:gates:z} \\
            \check{\phi}_r = \phi( \| W_r \quad U_r \, \check{x} \quad b_r \|_\infty).\label{eq:gru:exp_deltaiss:gates:r} 
        \end{align}
    \end{subequations}
\end{proposition}
\begin{pf}
    See Appendix \ref{proof:exp_deltaiss}.
\end{pf}

Knowing $\mu$ and $\lambda$, the NMPC law proposed in Section \ref{sec:mpc} can be synthesized by means of a suitable choice of the FHOCP's design parameters, i.e., the weight matrices ($Q$, $R$, $S$), the prediction horizon $N$, and the simulation horizon $M$, such that Theorem \ref{theorem:mpc_convergence} is fulfilled.

\subsection{GRU observer design} \label{sec:control:observer}
In the spirit of \cite{terzi2021lstm}, \cite{bonassi2021nonlinear}, to define a state estimate $\hat{x}_k$ with convergence guarantees, we propose to adopt a Luenberger-like structure that resembles that of the GRU model \eqref{eq:gru:model} whose state is to be observed.
Hence, the state observer candidate reads as
\begin{subequations} \label{eq:gru_observer:model}
	\begin{equation}\label{eq:mpc:gru_observer:model:state_space}
	\mathcal{O}(\Phi_o): \begin{dcases}
 	\hat{x}_{k+1} = \hat{z}_{k} \circ \hat{x}_{k} + (1 - \hat{z}_k) \circ \hat{r}_k \\
 	\hat{y}_k = U_o \hat{x}_k + b_o 
 \end{dcases},
\end{equation}
where $\hat{x}_k \in \mathbb{R}^{n_x}$ denotes the observer's state, whereas its gates $\hat{z}_k$ and $\hat{f}_k$, and the squashed input $\hat{r}_k$ are defined as
\begin{equation}\label{eq:mpc:gru_observer:model:gates}
	\begin{aligned}
		\hat{z}_k &= \sigma(W_z u_k + U_z \hat{x}_k + b_z + L_z(y_k - \hat{y}_k)), \\
		\hat{f}_k &= \sigma(W_f u_k + U_f \hat{x}_k + b_f + L_f(y_k - \hat{y}_k)), \\
		\hat{r}_k &= \phi(W_r u_k + U_f \hat{f}_k \circ \hat{x}_k + b_f).
	\end{aligned}
\end{equation}
\end{subequations}
Overall, the set of weights of the GRU observer \eqref{eq:gru_observer:model} is $\Phi_o = \Phi \cup \{ L_z, L_f \}$, where $\Phi$ is the set of fixed weights of the GRU system to be observed, while the observer's tuning parameters are the gains $L_{f}$ and $L_z$, which allow to improve the future state estimation $\hat{x}_{k+1}$ based on the known innovation, i.e., $y_k - \hat{y}_k$.
For compactness, the state observer \eqref{eq:gru_observer:model} may be denoted as
\begin{equation} \label{eq:gru_observer:generic}
	\mathcal{O}(\Phi_o) : \begin{dcases}
	   \hat{x}_{k+1} = \varphi_o(\hat{x}_k, u_k, y_k; \Phi_o) \\
		\hat{y}_k = g(\hat{x}_k; \Phi_o)
	\end{dcases}.
\end{equation}
It is here assumed that the initial state of the observer lies in $\mathcal{X}$.
This assumption, justified by the fact that such set is the invariant set of the system to be observed, allows to readily apply \cite[Lemma 2]{bonassi2021stability} to guarantee that $\mathcal{X}$ is also an invariant set of the state estimate $\hat{x}_k$.

\smallskip
\begin{theorem} \label{theorem:observer_convergence}
	Consider the observer gains $L_z$ and $L_f$. If there exists $\lambda_o \in (0, 1)$ such that, $\forall z \in [1 - \check{\sigma}_z, \check{\sigma}_z]$,
	\begin{subequations} \label{eq:gru_observer:condition}
	\begin{equation} \label{eq:gru_observer:condition:lambda_o}
		\kappa_o(z, L_z, L_f) < \lambda_o,
	\end{equation}
	where 
	\begin{equation} \label{eq:gru_observer:condition:kappa_o}
    \scalemath{0.9}{
	\begin{aligned}
		\kappa_o(z, L_z, L_f) =& z + (1- z)  \Big( \frac{1}{4} \check{x} \| U_f - L_f U_o \|_\infty + \check{\sigma}_f \Big) \| U_r \|_\infty \\
			&\quad +  \frac{1}{4} (\check{\phi}_r + \check{x}) \| U_z - L_z U_o \|_\infty
	\end{aligned}}
	\end{equation}
	\end{subequations}
	and $\check{\sigma}_z$, $\check{\sigma}_f$, and $\check{\phi}_r$ defined as in \eqref{eq:gru:exp_deltaiss:gates}, then \eqref{eq:gru_observer:model} is weak exponential observer.
\end{theorem}
\begin{pf}
	See Appendix \ref{proof:observer_convergence}.
\end{pf}

Note that, according to \eqref{eq:gru_observer:condition:lambda_o}, $\lambda_o$ provides a bound on the worst-case convergence rate of the observer.

Unfortunately, while Theorem \ref{theorem:observer_convergence} allows to certify the exponential convergence of the designed observer (given $L_z$ and $L_f$), it does not provide guidelines on how to tune these weights.
In the following we therefore recast the observer design problem as a convex optimization program, with the aim of computing the gains that attain the smallest possible worst-case convergence rate $\lambda_o$.

\begin{proposition}[Optimal observer tuning] \label{prop:gru_observer_design}
	The gains $L_z$ and $L_f$ of the state observer \eqref{eq:gru_observer:model} allowing to fulfill Theorem \ref{theorem:observer_convergence} while ensuring the fastest worst-case convergence rate $\lambda_o$ can be computed by solving the following convex optimization problem
	\begin{equation} \label{eq:gru_observer:design:problem}
	\begin{aligned}
		\lambda_o, L_z^\star, L_f^\star = \arg\min_{\tilde{\lambda}_o, L_z, L_f} \,\,& \tilde{\lambda}_o \\
		\text{s.t.} \quad & \kappa_{o}(\check{\sigma}_z, L_z, L_f) \leq \tilde{\lambda}_o \\
		&  \kappa_{o}(1-\check{\sigma}_z, L_z, L_f) \leq \tilde{\lambda}_o \\
		& 0 < \tilde{\lambda}_o < 1
	\end{aligned},
	\end{equation}
	where $\kappa_o(z, L_z, L_f)$ is defined as in \eqref{eq:gru_observer:condition:kappa_o}.
\end{proposition}
\begin{pf}
	See Appendix \ref{proof:gru_observer_design}.
\end{pf}

It is worth noticing that the exponential $\delta$ISS of the GRU model \eqref{eq:gru:model} entails the existence of a feasible solution to the optimization problem \eqref{eq:gru_observer:design:problem}.
Indeed, considering the suboptimal open-loop observer, corresponding to $L_z = L_f = 0_{n_x, n_y}$, one has that $\kappa_o(z, L_z, L_f) = \kappa(z)$, and hence $\lambda_o = \lambda \in (0, 1)$.
The open-loop observer thus represents a feasible solution to~\eqref{eq:gru_observer:design:problem}.

\section{Numerical results}\label{sec:casestudy}
\begin{figure}[t]
	\centering
	\includegraphics[width=0.6\columnwidth]{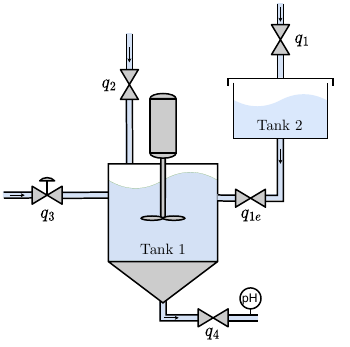}
    \vspace{-2mm}
	\caption{Schematic of the pH process benchmark system.}
	\label{fig:example:benchmark}
    \vspace{4mm}
    \includegraphics[width=0.8 \linewidth]{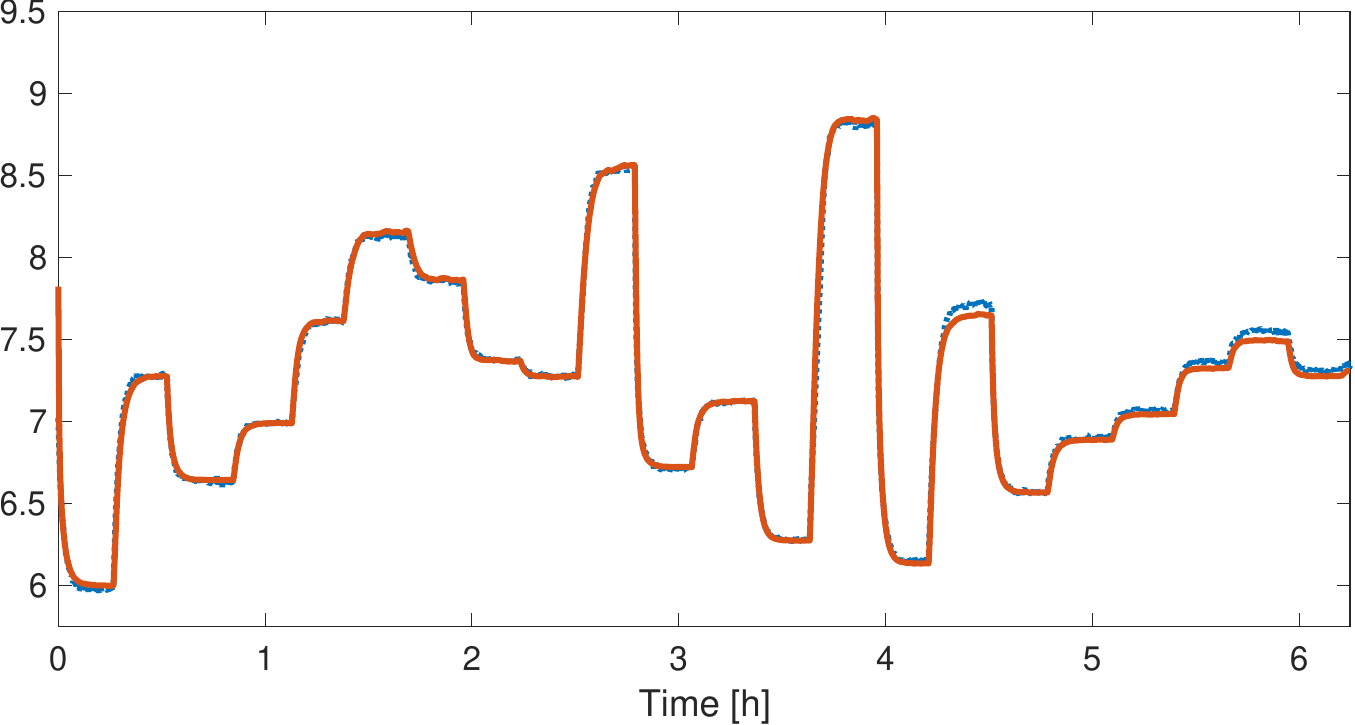}
    \caption{Open-loop simulation of the trained $\delta$ISS GRU model (red solid line) compared to the ground truth (blue dotted line), tested on an independent validation dataset.}
    \label{fig:example:model_testing}
\end{figure}

The proposed control framework is tested on the pH neutralization process described in \cite{henson1994adaptive}, and  depicted in Figure \ref{fig:example:benchmark}.
The system is composed of two tanks.
Tank $2$ is fed by the acid flowrate $q_1$, and its output is the flowrate $q_{1e}$.
The hydraulic dynamics of Tank $2$, being much faster than the others involved, are neglected, i.e. $q_1 = q_{1e}$.
Tank $1$ is fed by three flows, namely the acid flowrate $q_{1e}$, the buffer flowrate $q_2$, and the alkaline base flowrate $q_3$.
The terms $q_1$ and $q_2$ can not be manipulated and represent disturbances.
The alkaline flowrate $q_3$ can be modulated via a controllable valve, and thus represents the control variable. 
The output of the reactor tank is the fixed flowrate $q_4$, from which the pH is measured.
The control objective is to regulate the pH of the output flow to the (piecewise) constant setpoint.
The resulting model is a third-order nonlinear SISO system, whose equations and parameters are reported in \cite{henson1994adaptive}.

The system has been identified using a single-layer GRU model with $n_x = 7$ neurons according to the procedure illustrated in \cite{bonassi2021stability}, where the exponential $\delta$ISS of the network is also enforced. 
The dataset used to carry out the training procedure consists of approximately $25$ hours of operation, collected from a simulator of the system with a sampling time of $15$ seconds.
Note that the simulated output measurements have been corrupted by white noise.
The modeling performances of the trained GRU model have been tested on an independent validation dataset, showing the remarkable results depicted in Figure \ref{fig:example:model_testing}.

Then, a state observer in the form \eqref{eq:gru_observer:model} has been designed, by tuning the weights $L_z$ and $L_f$ as illustrated in Proposition \ref{prop:gru_observer_design}. 
In particular, the observer synthesis amounts to solving \eqref{eq:gru_observer:design:problem}, which has been performed using CVX, yielding $\lambda_o~=~0.93$.
The NMPC's FHOCP \eqref{eq:mpc:fhocp} has been formulated in accordance with Theorem~\ref{theorem:mpc_convergence}. 
The state weight matrix has been selected as $Q = I_{n_x, n_x}$, the input weight matrix as $R = 0.25$, and the terminal weight as $S = 2 Q$.
Note that the selected weights are positive definite and satisfy~\eqref{eq:mpc:sigmas_condition}.
Considering a prediction horizon $N = 20$, in order to choose the simulation horizon $M$ related to the terminal cost, the criterion \eqref{eq:mpc:horizon_condition} can be adopted, using  $\mu$ and $\lambda$ defined in \eqref{eq:gru:exp_deltaiss:mu_lambda}.
However, due to their conservativeness ($\lambda = 0.997$), an excessive simulation horizon would be required, i.e., $M \geq 440$.
We thus opted for numerically estimating $\lambda \in (0, 1)$ such that \eqref{eq:deltaiss:def} holds, with $\mu = \sqrt{7}$ fixed, for a sufficiently large number of pairs of state trajectories.
Such trajectories have been generated by simulating the model \eqref{eq:gru:model} with random pairs of initial states within $\mathcal{X}$ and random input sequences extracted from $\mathcal{U}_{0:T}$, with $T$ sufficiently high.
Note that this numerical approximation is made possible by the fact that the existence of such $\lambda$ is guaranteed by the model's exponential $\delta$ISS.
Considering $10^5$ trajectories\footnote{Note that this approach can be readily extended to MIMO systems, as long as the number of simulated trajectories is large enough to explore the set $\mathcal{U}$.} of length $T = 300$, the bound $\lambda \approx 0.9$ has been computed which, owing to \eqref{eq:mpc:horizon_condition}, implies $M \geq 15$.
The simulation horizon $M = 20$ has been therefore selected.

\begin{figure}
    \centering
    \includegraphics[width=0.8 \linewidth]{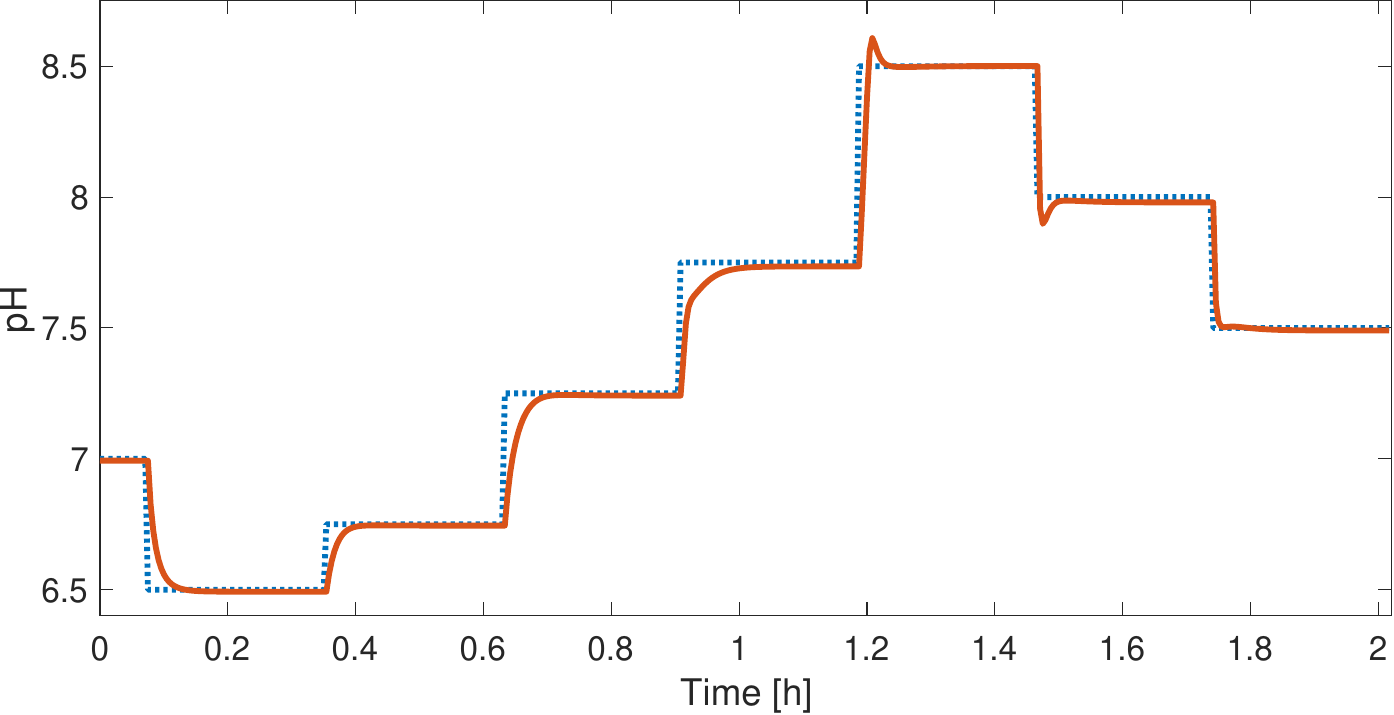}
    \caption{Closed-loop performances of the proposed control architecture: plant's output (red solid line) compared to the piecewise-constant reference signal (black dashed line).}
    \label{fig:example:cl_output}
    \vspace{5mm}
    \includegraphics[width=0.8 \linewidth]{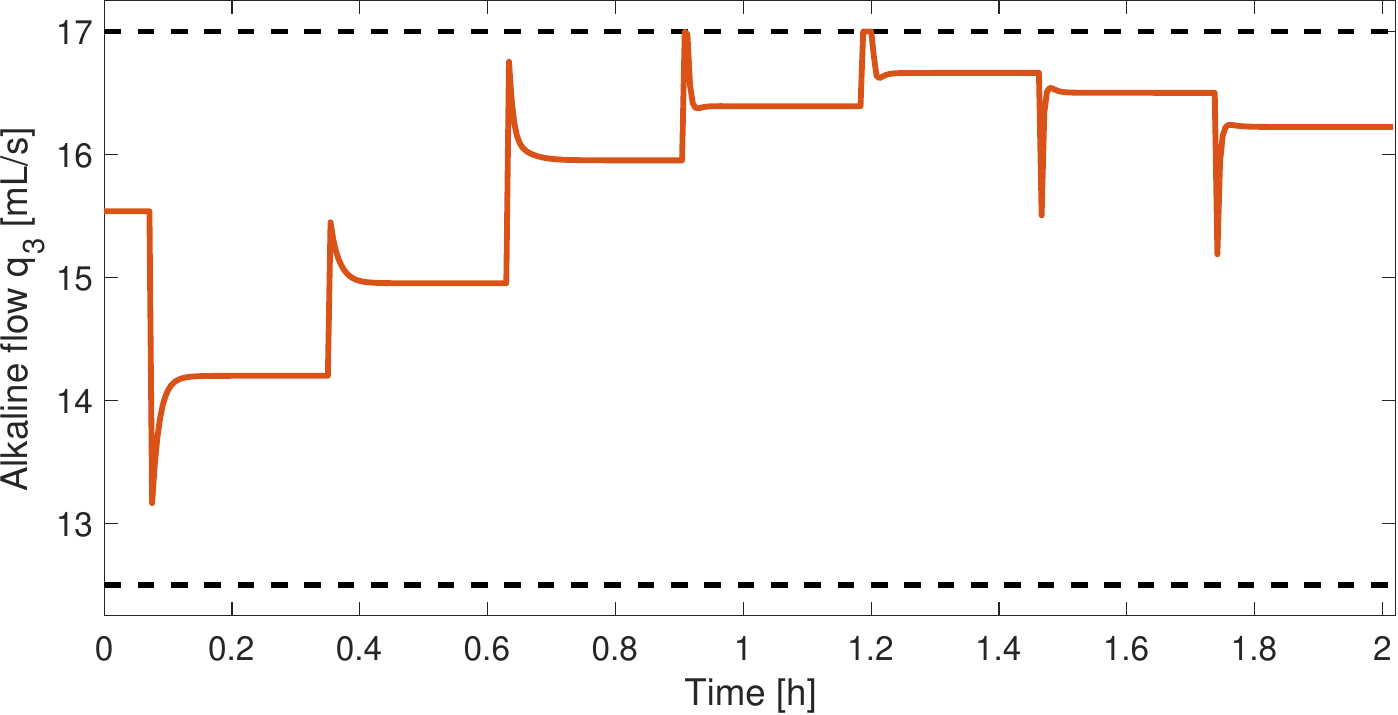}
    \caption{Control action requested by the proposed scheme (red solid line) and its saturation (black dashed line).}
    \label{fig:example:cl_input}
\end{figure}

The closed-loop performances of the designed control architecture have been tested using a piecewise-constant reference signal spacing the region of interest, considering the input constraint $\tilde{\mathcal{U}} = \mathcal{U}$.
Note that at every setpoint change, only the target equilibrium \eqref{ass:control:equilibrium} needs to be recomputed.
In Figure \ref{fig:example:cl_output}, the resulting closed-loop output trajectory is compared to the reference signal.
It is apparent that the proposed control architecture is able to accurately steer the output to the setpoint with negligible steady-state tracking error (in the range $\pm 0.02$ pH).
At the same time, the input saturation constraint is respected, see Figure \ref{fig:example:cl_input}.

Notably, the computational burden of the proposed architecture is rather low: the solution of the FHOCP \eqref{eq:mpc:fhocp} took $0.2 \pm 0.05$ seconds per timestep using CasADi, which is well below the sampling time.

\section{Conclusions}\label{sec:conclusion}
In this brief we presented an approach for designing nonlinear model predictive control laws with guaranteed closed-loop stability for Incrementally Input-to-State Stable ($\delta$ISS) systems.
In this framework, an explicit condition on the design of the terminal cost which ensures the attainment of the desired closed-loop properties was provided.
The proposed approach is particularly suitable for synthesis based on systems learned by Recurrent Neural Networks (RNN) models, in light of recent findings concerning their $\delta$ISS properties.
Therefore, we demonstrated how this control law can be synthesized for a popular  class of RNNs, i.e. Gated Recurrent Units. 
Future research efforts will be aimed at extending the proposed theoretical framework to include state constraints and offset-free output regulation, and at validating the proposed control strategy on experimental lab apparatuses.

\appendix
\section{Proofs}
\subsection{Proof of Theorem \ref{theorem:mpc_convergence}} \label{proof:mpc_convergence} 

Consider the optimal solution of \eqref{eq:mpc:fhocp} at time $k$. 
Let us denote the optimal control sequence as
	$u^{\star}_{k:k+N-1 \lvert k} = \{ u^{\star}_{k \lvert k} , ..., u^{\star}_{k+N-1 \lvert k} \}$,
and let $N_S = N + M$.
The corresponding optimal state trajectory, indicated by $x^{\star}_{k:k+N_S \lvert k} = \{ x^{\star}_{k \lvert k} , ..., x^{\star}_{k+N_S \lvert k} \}$,
is given by the evolution of the predictive model \eqref{eq:control:model}, initialized in $x_{k \lvert k} = \hat{x}_k$ (see \eqref{eq:mpc:fhocp:initialization}),  fed by 
$u^{\star}_{k:k+N-1 \lvert k}$ 
and, after the prediction horizon, by the constant, auxiliary control law $\bar{u}$.
Then, the optimal cost function $J_k^\star$ reads as
\begin{equation}\label{eq:proof:mpc:Jk_star}
\scalemath{0.875}{
	J_k^\star = \sum_{\tau=0}^{N-1} \Big( \| x_{k+\tau \lvert k}^\star - \bar{x} \|_Q ^2 + \| u_{k+\tau \lvert k}^\star - \bar{u} \|_R^2  \Big)  
	+ \sum_{\tau=N}^{N_S} \| x_{k+\tau \lvert k}^\star - \bar{x} \|_S ^2}.
\end{equation}
The goal of the proof is to show that $J_k^\star$ is a Lyapunov function for the closed-loop system.
To this end, we start by pointing out that
\begin{equation} \label{eq:proof:mpc:J_lb}
	J_k^\star \geq \| \hat{x}_k - \bar{x} \|_Q^2 \geq \ubar{\varsigma}_Q \| \hat{x}_k - \bar{x} \|_2^2.
\end{equation}
As customary in MPC literature \cite{rawlings2017model},  we then consider a sub-optimal (yet feasible) control sequence constantly equal to $\bar{u}$, i.e. $\tilde{u}_{k : k+N-1 \lvert k} = \{ \bar{u}, ..., \bar{u} \}$, and we denote by $\tilde{x}_{k : k+N_S \lvert k} = \{ \tilde{x}_{k \lvert k}, ..., \tilde{x}_{k+N_S \lvert k} \}$
the corresponding state trajectory of \eqref{eq:control:model}.
The suboptimality of $\tilde{u}_{k:k+N-1 \lvert k}$ and $\tilde{x}_{k : k+N_S \lvert k} $ entails that
\begin{equation} \label{eq:proof:mpc:J_ub}
\begin{aligned}
	J_k^\star & \leq \sum_{\tau=0}^{N-1} \| \tilde{x}_{k+\tau \lvert k} - \bar{x} \|_Q^2 +\sum_{\tau=N}^{N_S} \| \tilde{x}_{k+\tau \lvert k} - \bar{x} \|_S^2\; \\
 & \stackrel{\eqref{eq:mpc:sigmas_condition}}{\leq} \bar{\varsigma}_S \sum_{\tau=0}^{N_S} \| \tilde{x}_{k+\tau \lvert k} - \bar{x} \|_2^2 \stackrel{\eqref{eq:deltaiss:def}}{\leq}  \bar{\varsigma}_S \frac{\mu^2 }{1 - \lambda^2} \| \hat{x}_k - \bar{x} \|_2^.
\end{aligned}
\end{equation}
Combining \eqref{eq:proof:mpc:J_lb} and \eqref{eq:proof:mpc:J_ub} we get that the Lyapunov function candidate is bounded as 
\begin{equation*}
	\ubar{\varsigma}_Q \| \hat{x}_k - \bar{x} \|_2^2 \leq J_k^\star \leq \bar{\varsigma}_S \frac{\mu^2}{1 - \lambda^2} \| \hat{x}_k - \bar{x}_k \|_2^2.
\end{equation*}
At time $k+1$ the state observation $\hat{x}_{k+1}$ is available, and the optimization problem \eqref{eq:mpc:fhocp} is solved, yielding the optimal control sequence $u^{\star}_{k+1:k+N \lvert k+1} = \{ u^{\star}_{k+1 \lvert k+1} , ..., u^{\star}_{k+N \lvert k +1} \}$
and the optimal state trajectory $
	{x}_{k+1 : k+N_S+1 \lvert k+1}^\star = \{ {x}^\star_{k + 1\lvert k + 1}, ..., x_{k+N_S +1 \lvert k +1}^\star \}$.
The corresponding optimal cost function is denoted by $J^\star_{k+1}$.
Let us notice that the optimal control sequence computed at the previous NMPC iteration can be adopted as a suboptimal solution.
Indeed, let
\begin{subequations} \label{eq:proof:mpc:kp1_suboptimal}
\begin{equation}
	u_{k+1:k+N \lvert k + 1} = \{ u^{\star}_{k+1 \lvert k} , ..., u^{\star}_{k+N-1 \lvert k}, \bar{u} \}.
\end{equation}
The corresponding suboptimal state trajectory is
\begin{equation}
	x_{k+1:k+N+1 \lvert k} = \{ x_{k+1 \lvert k +1} , ..., x_{k+N \lvert k + 1}, x_{k+N+1 \lvert k + 1} \},
\end{equation}
\end{subequations}
where it is worth stressing that $x_{k+1 \lvert k+1} = \hat{x}_{k+1} \neq x^\star_{k+1 \lvert k}$.
Owing to the suboptimality of \eqref{eq:proof:mpc:kp1_suboptimal}, $J_{k+1}^\star$ can be bounded as
\begin{equation}\label{eq:proof:mpc:kp1_bound1}
\scalemath{1}{
	\begin{aligned}
		J_{k+1}^\star & \leq \sum_{\tau=1}^{N - 1} \big( \| x_{k+\tau \lvert k+1} - \bar{x} \|_Q^2  + \| u_{k+\tau \lvert k}^\star - \bar{u} \|_R^2  \big) \\
		& \quad + \| x_{k+N \lvert k+1} - \bar{x} \|_Q^2 + \sum_{\tau=N+1}^{N_S+1} \| x_{k+\tau \lvert k+1} - \bar{x} \|_S^2
	\end{aligned}}
\end{equation} 
Letting for consistency $x^\star_{k+N_S+1 \lvert k} = \varphi(x^\star_{k+N_S \lvert k}, \bar{u})$,  the following variable is introduced,  $\forall \tau \in \{1, ..., N_S + 1 \}$,
\begin{equation}
	e_{k+\tau|k+1} = x_{k+\tau \lvert k+1} - x^\star_{k+\tau \lvert k}.
\end{equation}
The bound \eqref{eq:proof:mpc:kp1_bound1} can be thus rewritten as
\begin{equation}\label{eq:proof:mpc:kp1_bound2}
\scalemath{0.9}{
	\begin{aligned}
		J_{k+1}^\star &\leq \sum_{\tau=1}^{N} \big\| \big( x_{k+\tau \lvert k}^\star - \bar{x}\big) + e_{k+\tau \lvert k+1} \big\|_Q^2  +  \sum_{\tau=1}^{N-1} \big\| u_{k+\tau \lvert k}^\star - \bar{u} \big\|_R^2   \\
		& \qquad + \big\| \big( x_{k+N +1 \lvert k}^\star - \bar{x} \big) + e_{k+N \lvert k+1} \big\|_Q^2 \\
		& \qquad + \sum_{\tau=N + 1}^{N_S + 1} \big\| \big( x_{k+ \tau \lvert k}^\star - \bar{x} \big) + e_{k+\tau \lvert k+1} \big\|_S^2 
	\end{aligned}}
\end{equation} 

In light of \eqref{eq:proof:mpc:Jk_star} and \eqref{eq:proof:mpc:kp1_bound2}, it follows that
\begin{subequations} \label{eq:proof:mpc:Delta_Jk_int0}
\begin{equation}
	\begin{aligned}
		J_{k+1}^\star - J_{k}^\star &\leq - \| x^\star_{k \lvert k} - \bar{x} \|_Q^2 - \| u^\star_{k \lvert k} - \bar{u} \|_R^2 + \Delta J_{a} + \Delta J_{b},
	\end{aligned}
\end{equation}
where $\Delta J_{a}$ and $\Delta J_{b}$ read as
\begin{equation}
\scalemath{0.9}{
\begin{aligned}
	\Delta J_{a} =& \sum_{\tau = 1}^{N-1} \Big( \big\|  \big( x_{k+\tau \lvert k}^\star - \bar{x}\big) + e_{k+\tau \lvert k+1} \big\|_Q^2  - \| x^\star_{k+t \lvert k} - \bar{x} \|_Q^2\Big)  \\
	& + \sum_{\tau = N + 1}^{N_S} \Big( \big\|  \big( x_{k+\tau \lvert k}^\star - \bar{x}\big) + e_{k+\tau \lvert k+1} \big\|_S^2  - \| x^\star_{k+t \lvert k} - \bar{x} \|_S^2\Big)
\end{aligned}}
\end{equation}
\begin{equation}
\scalemath{0.9}{
\begin{aligned}
	\Delta J_{b} &= \big\| \big( x_{k+N \lvert k}^\star - \bar{x} \big) + e_{k+N \lvert k+1} \big\|_Q^2 - \|  x_{k+N \lvert k}^\star - \bar{x} \|_S^2 \\
	&\qquad + \big\| \big( x_{k+N_S+1 \lvert k}^\star - \bar{x} \big) + e_{k+N_S+1 \lvert k+1} \big\|_S^2.
\end{aligned}}
\end{equation}
\end{subequations}

Let us now derive a bound for the term $\Delta J_a$.
By noticing that $\| v + w \|_Q^2 = \| v \|_Q^2 + \| w \|_Q^2 + 2 v^\prime Q w$, it holds that
\begin{equation}\label{eq:proof:mpc:Delta_Ja}
\scalemath{0.9}{
\begin{aligned}
	\Delta J_{a} \leq& \sum_{t=1}^{N-1} \Big( \| e_{k+\tau \lvert k+1} \|_Q^2 + 2 (x_{k+\tau \lvert k}^\star - \bar{x})^\prime Q e_{k+\tau \lvert k+1}\Big) \\
	& + \sum_{t=N+1}^{N_S} \Big( \| e_{k+\tau \lvert k+1} \|_S^2 + 2 (x_{k+\tau \lvert k}^\star - \bar{x})^\prime S e_{k+\tau \lvert k+1}\Big).
\end{aligned}}
\end{equation}
Since $x_{k+t \lvert k}^\star$ belongs to the invariant set $\mathcal{X}$, there exist finite scalars $\mu_{a1} > 0$ and $\mu_{a2} > 0$ such that \eqref{eq:proof:mpc:Delta_Ja} can be upper bounded as
\begin{equation} \label{eq:proof:mpc:Delta_Ja_bound}
\begin{aligned}
	\Delta J_{a} &\leq \mu_{a1} \bar{\varsigma}_Q
	 \sum_{t=1}^{N - 1}  \big( \| e_{k+\tau \lvert k+1} \|_2^2 + \| e_{k+\tau \lvert k+1} \|_2 \big) \\
	 & \quad + \mu_{a2} \bar{\varsigma}_S \sum_{t=N + 1}^{N_S}  \big( \| e_{k+\tau \lvert k+1} \|_2^2 + \| e_{k+\tau \lvert k+1} \|_2 \big).
\end{aligned}
\end{equation}
Analogously, the term $\Delta J_b$ can be bounded as
\begin{equation} \label{eq:proof:mpc:Delta_Jb_bound}
\scalemath{0.9}{
\begin{aligned}
	\Delta J_{b} \leq&  \|  x_{k+N \lvert k}^\star - \bar{x} \|_Q^2 - \|  x_{k+N \lvert k}^\star - \bar{x} \|_S^2 + \|  x_{k+ N_S + 1 \lvert k}^\star - \bar{x} \|_S^2 \\
	& + \| e_{k+N \lvert k+1} \|_Q^2 + \| e_{k+N_S + 1 \lvert k+1} \|_S^2 \\
	& + 2 (x_{k+N \lvert k}^\star - \bar{x})^\prime Q e_{k+N \lvert k+1} \\
	& +  2 (x_{k+N_S + 1 \lvert k}^\star - \bar{x})^\prime S e_{k+N_S + 1 \lvert k+1} 
\end{aligned}}
\end{equation}
We now show that if \eqref{eq:mpc:as_conditions} holds, then 
\begin{equation}\label{eq:proof:mpc:Delta_Jb_Step1bis}
	\|  x_{k+ N_S + 1 \lvert k}^\star - \bar{x} \|_S^2 < \|  x_{k+N \lvert k}^\star - \bar{x} \|_{S - Q}^2,
\end{equation}
where $S - Q \succ 0$ due to \eqref{eq:mpc:sigmas_condition}.
To this end, let us point out that if
\begin{equation}\label{eq:proof:mpc:Delta_Jb_Step2}
	\bar{\varsigma}_S \|  x_{k+ N_S + 1 \lvert k}^\star - \bar{x} \|_2^2 < (\ubar{\varsigma}_S - \bar{\varsigma}_Q) \|  x_{k+N \lvert k}^\star - \bar{x} \|_{2}^2,
\end{equation}
then \eqref{eq:proof:mpc:Delta_Jb_Step1bis} surely holds.
Since for $\tau \geq N$ the constant, auxiliary control law $\bar{u}$ is applied, it holds that 
\begin{equation}
	\bar{\varsigma}_S  \|  x_{k+ N_S + 1 \lvert k}^\star - \bar{x} \|_2^2 \stackrel{\eqref{eq:deltaiss:def}}{\leq} \bar{\varsigma}_S  \mu^2  \lambda^{2(M + 1)} \|  x_{k+ N \lvert k}^\star - \bar{x} \|_2^2.
\end{equation}
Under \eqref{eq:mpc:horizon_condition} one can thus guarantee that
\begin{equation}\label{eq:proof:mpc:Delta_Jb_Step3}
	\scalemath{0.9}{
	 \bar{\varsigma}_S  \mu^2  \lambda^{2(M + 1)} \|  x_{k+ N \lvert k}^\star - \bar{x} \|_2^2 \leq (\ubar{\varsigma}_S - \bar{\varsigma}_Q) \|  x_{k+N \lvert k}^\star - \bar{x} \|_{2}^2.}
\end{equation}
By means of a chain of inequalities, this entails that \eqref{eq:proof:mpc:Delta_Jb_Step2} holds.

Noting that $\mathcal{X}$ is an invariant set, $x_{k+N \lvert k} \in \mathcal{X}$ holds. 
Therefore, there exist finite scalars $\mu_{b1} > 0$ and $\mu_{b 2} > 0$ such that \eqref{eq:proof:mpc:Delta_Jb_bound} can be upper bounded by
\begin{equation} \label{eq:proof:mpc:Delta_Jb_bound_upper}
\begin{aligned}
	\Delta J_{b} &\leq \mu_{b1} \bar{\varsigma}_Q  \big(  \| e_{k + N \lvert k+1} \|_2^2 + \| e_{k + N \lvert k+1} \|_2 \big) \\
	 & \quad + \mu_{b2} \bar{\varsigma}_S \big( \| e_{k+N_S+1 \lvert k+1} \|_2^2 + \| e_{k+N_S+1 \lvert k+1} \|_2 \big).
\end{aligned}
\end{equation}

By applying the bounds retrieved in \eqref{eq:proof:mpc:Delta_Ja_bound} and \eqref{eq:proof:mpc:Delta_Jb_bound_upper} to  \eqref{eq:proof:mpc:Delta_Jk_int0}, and recalling \eqref{eq:mpc:sigmas_condition}, one can guarantee the existence of a finite $\mu_e > 0$ such that
\begin{equation} \label{eq:proof:convergence:Delta_J_bound}
\begin{aligned}	J_{k+1}^\star - J_{k}^\star \leq& - \| x^\star_{k \lvert k} - \bar{x} \|_Q^2 \\
	&+ \underbrace{\mu_e \sum_{\tau=1}^{N_S+1} \big(  \| e_{k + \tau \lvert k+1} \|_2^2 + \| e_{k + \tau \lvert k+1} \|_2 \big)}_{\varrho_{e, k}}.
\end{aligned}
\end{equation}
To conclude the proof, we show that the term $\varrho_{e, k}$ exponentially converges to zero with the time index $k$.
To this end, let us point out that, by definition,
\begin{equation} 
\begin{aligned}
	e_{k+1 \lvert k+1} &= x_{k+1 \lvert k+1} - x_{k+1 \lvert k}^\star \\
	&= \varphi_o(\hat{x}_k, u_{k \lvert k}^\star, y_k) - \varphi(\hat{x}, u_{k \lvert k}^\star),
\end{aligned}
\end{equation}
Being the observer a weak exponential one, it holds that
\begin{equation} \label{eq:proof:convergence:varrho_1_bound}
	\| e_{k+1 \lvert k+1} \|_2^2 \stackrel{\eqref{eq:control:observer_convergence}}{\leq} \mu_o^2 \tilde{\lambda}_o^2 \| \hat{x}_k - x_k \|_2^2.
\end{equation}
Let now, for the sake of compactness,
\begin{equation*}
	u^\star_{k+1:k+N_S \lvert k} = \{ u^\star_{k+1 \lvert k}, ..., u^\star_{k+N \lvert k}, \bar{u}, ..., \bar{u} \}.
\end{equation*}
For any $t \in \{ 2, ..., N_S + 1\}$ by definition we have that
\begin{equation}
	e_{k+t \lvert k+1} = x_{k+t \lvert k+1} - x_{k+t \lvert k}^\star, 
\end{equation}
where $	x_{k+t \lvert k+1} = x_{k+t}(x_{k+1 \lvert k+1}, u_{k+1:k+t \lvert k}^\star)$ and $x_{k+t \lvert k}^\star = x_{k+t}(x_{k+1 \lvert k}^\star, u_{k+1:k+t \lvert k}^\star)$.

In view of the model's exponential $\delta$ISS, the term $\| e_{k+t \lvert k+1} \|_2$ can be bounded as
\begin{equation}\label{eq:proof:convergence:varrho_2_bound}
\begin{aligned}
		\| e_{k+t \lvert k+1} \|_2^2 &\stackrel{\eqref{eq:deltaiss:def}}{\leq} \mu^2 \lambda^{2(t-1)} \| x_{k+1 \lvert k+1} -x_{k+1 \lvert k}^\star \|_2^2 \\
		& \stackrel{\eqref{eq:proof:convergence:varrho_1_bound}}{\leq} \mu^2 \mu_o^2 \, \tilde{\lambda}_o^{2} \lambda^{2(t-1)} \| \hat{x}_k - x_k \|_2^2.
\end{aligned}
\end{equation}
Therefore, owing to \eqref{eq:proof:convergence:varrho_1_bound} and \eqref{eq:proof:convergence:varrho_2_bound}, and since the observer is exponentially converging, there exists $\mu_{\varrho} > 0$ such that
\begin{equation}
	\varrho_{e, k} \leq \mu_{\varrho} \| \hat{x}_k - x_k \|_2^2 \leq \mu_{\varrho} \tilde{\lambda}_o^{2k} \| \hat{x}_0 - x_0 \|_2^2.
\end{equation}
That is, the perturbation term $\varrho_{e, k}$  of \eqref{eq:proof:convergence:Delta_J_bound} exponentially converges to zero.  
The nominal closed-loop asymptotic stability can be therefore proven following \cite{scokaert1997discrete}. \hfill\qed

\subsection{Proof of Proposition \ref{prop:exp_deltaiss}}\label{proof:exp_deltaiss}
Consider two state trajectories, $x_{a, 0:k}$ and $x_{b, 0:k}$, yield by the initial states $x_{a, 0}$ and $x_{b, 0}$, and by the input sequences  $u_{a, 0:k}$ and $u_{b, 0:k}$.
If the GRU is exponentially $\delta$ISS by  \cite[Theorem 2]{bonassi2021stability}, in view of \cite[(A.23)-(A.24)]{bonassi2021stability} the following inequality holds 
\begin{equation}
\scalemath{0.85}{
	\| x_{a, k+1} - x_{b, k+1} \|_\infty \leq \kappa_x(z) \| x_{a, k} - x_{b, k} \|_\infty + \check{\kappa}_u \| u_{a, k} - u_{b, k} \|_\infty,}
\end{equation}
for any time instant $k$ and any scalar $z \in [1 - \check{\sigma}_z, \check{\sigma}_z ]$, where $\check{\sigma}_z$ is defined in \eqref{eq:gru:exp_deltaiss:gates:z} and $\kappa_x(z)$ reads as  \eqref{eq:gru:exp_deltaiss:kappa}, for a suitably defined $\check{\kappa}_u$.
This implies that, letting 
\begin{subequations}
\begin{equation} \label{eq:proof:deltaiss:lambda:lambda_def}
	\lambda = \max_{z \in [1 - \check{\sigma}_z, \check{\sigma}_z ]} \kappa_x(z),
\end{equation}
it must hold that
\begin{equation} \label{eq:proof:deltaiss:lambda:ineq}
	\| x_{a, k+1} - x_{b, k+1} \|_\infty \leq \lambda  \| x_{a, k} - x_{b, k} \|_\infty + \check{\kappa}_u \| u_{a, k} - u_{b, k} \|_\infty.
\end{equation}
\end{subequations}
Let us now point out that $\kappa_x(z)$ consists of a constant term, i.e. $\frac{1}{4} (\check{\phi}_r + \check{x}) \| U_z \|_\infty$, plus the convex combination, weighted by $z \in [1 - \check{\sigma}_z, \check{\sigma}_z ] \subset (0, 1)$, of two positive terms.
Therefore, the absolute maximum of $\kappa_x(z)$ happens at the closed boundaries of $z$, which yields \eqref{eq:gru:exp_deltaiss:mu_lambda:lambda}, where $\lambda \in (0, 1)$ is guaranteed by \cite[Theorem 2]{bonassi2021stability}.

Then, by iterating \eqref{eq:proof:deltaiss:lambda:ineq}, one gets
\begin{equation}\label{eq:proof:deltaiss:ineq_inf}
	\| x_{a, k} - x_{b, k} \|_\infty \leq \lambda^k \| x_{a, 0} - x_{b, 0} \|_\infty + \frac{\check{\kappa}_u}{1-\lambda} \| u_{a, k} - u_{b, k} \|_\infty.
\end{equation}
Since for any $v \in \mathbb{R}^n$ it holds that $\frac{1}{\sqrt{n}} \| v \|_2 \leq \| v \|_\infty \leq \| v \|_2$,
inequality \eqref{eq:proof:deltaiss:ineq_inf} implies
\begin{equation}\label{eq:proof:deltaiss:ineq_2}
\begin{aligned}
	\| x_{a, k} - x_{b, k} \|_2 \leq& \sqrt{n_x} \lambda^k \| x_{a, 0} - x_{b, 0} \|_\infty \\
	&+ \sqrt{n_x}  \frac{\check{\kappa}_u}{1-\lambda} \| u_{a, k} - u_{b, k} \|_\infty,
\end{aligned}
\end{equation}
i.e., $\mu = \sqrt{n_x}$. \hfill \qed

\subsection{Proof of Theorem \ref{theorem:observer_convergence}} \label{proof:observer_convergence}
Let $x_k(x_0, u_{0:k})$ be the state of the system \eqref{eq:gru:model}, and $\hat{x}_k(\hat{x}_0, u_{0:k}, y_{0:k})$ the state estimate yield by the observer \eqref{eq:gru_observer:model}, where $y_\tau = g(x_\tau)$.
In the following, these trajectories are compactly denoted by $x_k$ and $\hat{x}_k$, respectively.

Consider the $j$-th component of the state observation error at the generic step $k+1$. Summing and subtracting the terms $[z_k]_j [\hat{x}_k]_j$ and $(1 - [z_k]_j) [ \hat{r}_k ]_j $ we get
\begin{equation} \label{eq:proof:observer:error}
\scalemath{0.9}{
\begin{aligned}
	[x_{k+1}]_j - [\hat{x}_{k+1}]_j =& [z_k]_j \big([x_k]_j - [\hat{x}_k]_j \big) + \big( [z_k]_j - [\hat{z}_k]_j \big) [\hat{x}_k]_j \\
	&+ (1 - [z_k]_j) \big( [ r_k ]_j - [ \hat{r}_k ]_j \big) \\
    &+ \big([z_k]_j - [\hat{z}_k]_j \big) [ \hat{r}_k ]_j 	
\end{aligned}}
\end{equation}
Along the lines of the proof of \cite[Theorem 2]{bonassi2021stability}, we take the absolute value of both sides of \eqref{eq:proof:observer:error}.
In light of the boundedness of the sigmoidal activation function, it holds that $[z_k]_j \in (0, 1)$ and $[\hat{z}_k]_j \in (0, 1)$, which leads to
\begin{equation} \label{eq:proof:observer:error_ineq}
    \scalemath{0.85}{
	\begin{aligned}
		\big\lvert [x_{k+1}]_j - [\hat{x}_{k+1}]_j \big\lvert \leq & [z_k]_j \big\lvert [x_k]_j - [\hat{x}_k]_j \big\lvert + \big\lvert [z_k]_j - [\hat{z}_k]_j \big\lvert \, \lvert [\hat{x}_k]_j \lvert \\
		& + (1 - [z_k]_j) \big\lvert [ r_k ]_j - [ \hat{r}_k ]_j \big\lvert\\
		&+ \big\lvert [z_k]_j - [\hat{z}_k]_j \big\lvert \, \lvert [ \hat{r}_k ]_j \lvert 	
	\end{aligned}}
\end{equation}

\begin{subequations} \label{eq:proof:observer:bounds}
Since following \cite[Lemma 1]{bonassi2021stability} one can easily prove that  $\mathcal{X}$ is an invariant set for $\hat{x}$, it holds that
\begin{equation}  \label{eq:proof:observer:bounds:x}
	\lvert [ \hat{x}_k]_j \lvert \leq \| \hat{x}_k \|_\infty \leq \check{x}.
\end{equation}
Moreover, since $\phi$ is strictly increasing and Lipschitz continuous, in view of Assumption \ref{ass:bounded_input}, it holds that
\begin{equation}
\scalemath{0.975}{
\begin{aligned}
	\lvert [ \hat{r}_k ]_j \lvert &\leq \| \hat{r}_k \|_\infty \leq \max_{u \in \mathcal{U}} \big\| \phi(W_r u_k + U_r \hat{x}_k + b_r) \big\|_\infty \stackrel{\eqref{eq:gru:exp_deltaiss:gates:r}}{:=} \check{\phi}_r.
\end{aligned}}
\end{equation}
Exploit the $\frac{1}{4}$-Lipschitzianity of $\sigma$ and the linearity of output transformation $g(x_k)$, 
\begin{equation}
\scalemath{0.9}{
\begin{aligned}
	\big\lvert [z_k]_j - [\hat{z}_k]_j \big\lvert &\leq \| z_k - \hat{z}_k \|_\infty \leq  \frac{1}{4} \| U_z - L_z U_o \|_\infty \| x_k - \hat{x}_k \|_\infty, \\
	\big\lvert [f_k]_j - [\hat{f}_k]_j \big\lvert &\leq \| f_k - \hat{f}_k \|_\infty \leq  \frac{1}{4} \| U_f - L_f U_o \|_\infty \| x_k - \hat{x}_k \|_\infty.
	\end{aligned}}
\end{equation}
Since $\phi$ is $1$-Lipschitz, the following chain of inequalities also holds true
\begin{equation}
\scalemath{0.9}{
\begin{aligned}
	\big\lvert [r_k]_j - [\hat{r}_k]_j \big\lvert &\leq \| r_k - \hat{r}_k \|_\infty \\
	& \leq \| U_r \|_\infty \big\| (f_k - \hat{f}_k) \circ \hat{x}_k + f_k \circ (x_k - \hat{x}_k) \big\|_\infty \\
	& \leq \| U_r \|_\infty \Big[ \check{x} \| f_k - \hat{f}_k \|_\infty  + \check{\sigma}_f \| x_k - \hat{x}_k \|_\infty \Big] \\
	& \leq \| U_r \|_\infty \Big( \frac{1}{4} \check{x} \| U_f - L_f U_o \|_\infty + \check{\sigma}_f \Big) \| x_k - \hat{x}_k \|_\infty.
\end{aligned}}
\end{equation}
\end{subequations}
Applying the bounds \eqref{eq:proof:observer:bounds}, from  \eqref{eq:proof:observer:error_ineq} one gets
\begin{equation} \label{eq:proof:observer:error_ineq2}
\begin{aligned}
	\big\lvert [x_{k+1}]_j - [\hat{x}_{k+1}]_j \big\lvert &\leq {\kappa}_{o}([z_k]_j, L_z, L_f) \| x_k - \hat{x}_k \|_\infty,
\end{aligned}
\end{equation}
where $\kappa_o = \kappa_o(\cdot, L_z, L_f)$ is defined as in \eqref{eq:gru_observer:condition:kappa_o}.
Letting 
\begin{equation}
	\check{\sigma}_z = \sigma( \| W_z \quad U_z \check{x} \quad b_z \|_\infty),
\end{equation}
it is easily proved that $[z_k]_j \in [ 1-\check{\sigma}_z, \check{\sigma}_z] \subset (0, 1)$, see \cite{bonassi2021stability}.
In force of \eqref{eq:gru_observer:condition:lambda_o}, \eqref{eq:proof:observer:error_ineq2} is entailed by
\begin{equation} \label{eq:proof:observer:error_bound}
	\| x_{k+1} - \hat{x}_{k+1} \|_\infty \leq \lambda_o \| x_{k} - \hat{x}_k \|_\infty.
\end{equation}
Iterating \eqref{eq:proof:observer:error_bound} in time, we finally get
\begin{equation} \label{eq:proof:observer:error_bound_iterated}
	\| x_{k} - \hat{x}_{k} \|_\infty \leq \lambda_o^k \| x_{0} - \hat{x}_0 \|_\infty,
\end{equation}
which, by standard norm arguments, implies the observer's exponential convergence, in the sense specified by Definition~\ref{def:observer_convergence}, with $\mu_o = \sqrt{n_x}$. \hfill\qed
\subsection{Proof of Proposition \ref{prop:gru_observer_design}}\label{proof:gru_observer_design}
First, let us point out that, as evident from \eqref{eq:proof:observer:error_bound}, $\lambda_o$ represents a bound on the observer's worst-case convergence rate.
Therefore, the ``optimal'' gains of the observer are those that entail the smallest possible $\lambda_o$.
The observer design problem is therefore set up as a min-max optimization problem, where the observer gains corresponding to the smallest $\lambda_o$ are retrieved
\begin{equation}  \label{eq:proof:observer_design:min_max}
	\lambda_o = \min_{L_z, L_f} \bigg\{ \max_{z \in [ 1 - \check{\sigma}_z, \check{\sigma}_z]} \kappa_o(z, L_z, L_f) \bigg\}.
\end{equation}
Notice that the $\delta$ISS of the underlying GRU model implies that the optimal solution of \eqref{eq:proof:observer_design:min_max} does indeed satisfy Theorem~\ref{theorem:observer_convergence}, i.e. $\tilde{\lambda}_o \in (0, 1)$.
In fact, by taking the suboptimal gains $L_z = 0_{n_c, n_y}$ and $L_f = 0_{n_c, n_y}$ and recalling \eqref{eq:proof:deltaiss:lambda:lambda_def}, it holds that
	$\kappa_o(z, L_z, L_f) = \kappa_x(z) \leq \lambda < 1,\,$
where $\kappa_x(z)$ is that defined in \eqref{eq:gru:exp_deltaiss:kappa}.
By definition the optimal solution of \eqref{eq:proof:observer_design:min_max} thus $\lambda_{o} \leq \lambda_\delta < 1$, i.e., the optimal gains $L_z^\star$ and $L_f^\star$ satisfy Theorem \ref{theorem:observer_convergence}.
Because $\frac{\partial \kappa_o}{\partial z}$ does not depend on $z$, it holds that
	\begin{equation}
    \scalemath{0.9}{
	\begin{aligned}
		&\max_{z \in [ 1 - \check{\sigma}_z, \check{\sigma}_z]} \kappa_o(z, L_z, L_f) \\
		&\qquad= \max\big( \kappa_o(\check{\sigma}_z, L_z, L_f), \kappa_o(1-\check{\sigma}_z, L_z, L_f) \big).	
	\end{aligned}}
	\end{equation}
	The optimization problem \eqref{eq:proof:observer_design:min_max} can be therefore recast in the convex optimization problem \eqref{eq:gru_observer:design:problem}. \hfill\qed

\section*{Acknowledgments}
This project has received funding from the European Union’s Horizon 2020 research and innovation programme under the Marie Skłodowska-Curie grant agreement No. 953348. 

\smallskip
The work of Riccardo Scattolini is carried out within the MICS (Made in Italy – Circular and Sustainable) Extended Partnership and received funding from Next-Generation EU (Italian PNRR – M4 C2, Invest 1.3 – D.D. 1551.11-10-2022, PE00000004). CUP MICS D43C22003120001

\bibliographystyle{plain}
\bibliography{Bibliografia}

\begin{thebibliography}{10}

\bibitem{goodfellow2016deep}
Yoshua Bengio, Ian Goodfellow, and Aaron Courville.
\newblock {\em Deep learning}, volume~1.
\newblock MIT press Massachusetts, USA, 2017.

\bibitem{boccia2014stability}
Andrea Boccia, Lars Gr{\"u}ne, and Karl Worthmann.
\newblock Stability and feasibility of state constrained {MPC} without stabilizing terminal constraints.
\newblock {\em Systems \& control letters}, 72:14--21, 2014.

\bibitem{bonassi2021stability}
Fabio Bonassi, Marcello Farina, and Riccardo Scattolini.
\newblock On the stability properties of gated recurrent units neural networks.
\newblock {\em Systems \& Control Letters}, 157:105049, 2021.

\bibitem{bonassi2022offset}
Fabio Bonassi, Marcello Farina, Jing Xie, and Riccardo Scattolini.
\newblock {A}n {O}ffset-{F}ree {N}onlinear {MPC} scheme for systems learned by {N}eural {NARX} models.
\newblock In {\em 61st IEEE Conference on Decision and Control (CDC)}, 2022.

\bibitem{bonassi2022survey}
Fabio Bonassi, Marcello Farina, Jing Xie, and Riccardo Scattolini.
\newblock {O}n {R}ecurrent {N}eural {N}etworks for learning-based control: recent results and ideas for future developments.
\newblock {\em Journal of Process Control}, 114:92--104, 2022.

\bibitem{bonassi2021nonlinear}
Fabio Bonassi, Caio~Fabio Oliveira~da Silva, and Riccardo Scattolini.
\newblock Nonlinear {MPC} for {O}ffset-{F}ree {T}racking of systems learned by {GRU} {N}eural {N}etworks.
\newblock In {\em 3rd IFAC Conference on Modelling, Identification and Control of Nonlinear Systems (MICNON 2021)}, 2021.

\bibitem{armenio2019echo}
Luca {Bugliari Armenio}, Enrico Terzi, Marcello Farina, and Riccardo Scattolini.
\newblock {E}cho {S}tate {N}etworks: analysis, training and predictive control.
\newblock In {\em 2019 18th European Control Conference (ECC)}, pages 799--804. IEEE, 2019.

\bibitem{armenio2019model}
Luca {Bugliari Armenio}, Enrico Terzi, Marcello Farina, and Riccardo Scattolini.
\newblock {M}odel {P}redictive {C}ontrol {D}esign for {D}ynamical {S}ystems {L}earned by {E}cho {S}tate {N}etworks.
\newblock {\em IEEE Control Systems Letters}, 3(4):1044--1049, 2019.

\bibitem{henson1994adaptive}
Michael~A Henson and Dale~E Seborg.
\newblock Adaptive nonlinear control of a ph neutralization process.
\newblock {\em IEEE transactions on control systems technology}, 2(3):169--182, 1994.

\bibitem{hewing2020learning}
Lukas Hewing, Kim~P Wabersich, Marcel Menner, and Melanie~N Zeilinger.
\newblock Learning-based model predictive control: Toward safe learning in control.
\newblock {\em Annual Review of Control, Robotics, and Autonomous Systems}, 3:269--296, 2020.

\bibitem{kohler2019nonlinear}
Johannes K{\"o}hler, Matthias~A M{\"u}ller, and Frank Allg{\"o}wer.
\newblock A nonlinear model predictive control framework using reference generic terminal ingredients.
\newblock {\em IEEE Transactions on Automatic Control}, 65(8):3576--3583, 2019.

\bibitem{lanzetti2019recurrent}
Nicolas Lanzetti, Ying~Zhao Lian, Andrea Cortinovis, Luis Dominguez, Mehmet Mercang{\"o}z, and Colin Jones.
\newblock Recurrent neural network based {MPC} for process industries.
\newblock In {\em 2019 18th European Control Conference (ECC)}, pages 1005--1010. IEEE, 2019.

\bibitem{magni2004stabilization}
L~Magni, G~De~Nicolao, and Riccardo Scattolini.
\newblock On the stabilization of nonlinear discrete-time systems with output feedback.
\newblock {\em International Journal of Robust and Nonlinear Control: IFAC-Affiliated Journal}, 14(17):1379--1391, 2004.

\bibitem{magni2001stabilizing}
Lalo Magni, Giuseppe De~Nicolao, Lorenza Magnani, and Riccardo Scattolini.
\newblock A stabilizing model-based predictive control algorithm for nonlinear systems.
\newblock {\em Automatica}, 37(9):1351--1362, 2001.

\bibitem{mayne2000constrained}
David~Q Mayne, James~B Rawlings, Christopher~V Rao, and Pierre~OM Scokaert.
\newblock Constrained model predictive control: Stability and optimality.
\newblock {\em Automatica}, 36(6):789--814, 2000.

\bibitem{mohajerin2019multistep}
Nima Mohajerin and Steven~L Waslander.
\newblock Multistep prediction of dynamic systems with recurrent neural networks.
\newblock {\em IEEE transactions on neural networks and learning systems}, 30(11):3370--3383, 2019.

\bibitem{patan2014neural}
Krzysztof Patan.
\newblock Neural network-based model predictive control: Fault tolerance and stability.
\newblock {\em IEEE Transactions on Control Systems Technology}, 23(3):1147--1155, 2014.

\bibitem{rawlings2017model}
James~Blake Rawlings, David~Q Mayne, and Moritz Diehl.
\newblock {\em Model predictive control: theory, computation, and design}, volume~2.
\newblock Nob Hill Publishing Madison, WI, 2017.

\bibitem{scokaert1997discrete}
Pierre~OM Scokaert, James~B Rawlings, and Edward~S Meadows.
\newblock Discrete-time stability with perturbations: Application to model predictive control.
\newblock {\em Automatica}, 33(3):463--470, 1997.

\bibitem{soloperto2022nonlinear}
Raffaele Soloperto, Johannes Koehler, and Frank Allgower.
\newblock A nonlinear mpc scheme for output tracking without terminal ingredients.
\newblock {\em IEEE Transactions on Automatic Control}, 2022.

\bibitem{terzi2021lstm}
Enrico Terzi, Fabio Bonassi, Marcello Farina, and Riccardo Scattolini.
\newblock Learning model predictive control with long short-term memory networks.
\newblock {\em International Journal of Robust and Nonlinear Control}, 31(18):8877--8896, 2021.

\bibitem{wu2020process}
Zhe Wu, David Rincon, and Panagiotis~D Christofides.
\newblock Process structure-based recurrent neural network modeling for model predictive control of nonlinear processes.
\newblock {\em Journal of Process Control}, 89:74--84, 2020.

\bibitem{wu2019machine}
Zhe Wu, Anh Tran, David Rincon, and Panagiotis~D Christofides.
\newblock Machine learning-based predictive control of nonlinear processes. part i: Theory.
\newblock {\em AIChE Journal}, 65(11), 2019.

\bibitem{zarzycki2021lstm}
Krzysztof Zarzycki and Maciej {\L}awry{\'n}czuk.
\newblock Lstm and gru neural networks as models of dynamical processes used in predictive control: A comparison of models developed for two chemical reactors.
\newblock {\em Sensors}, 21(16):5625, 2021.

\end{thebibliography}

\end{document}